\def\t{e^{{{(r^2 + \frac{\beta^2}{\pi^2}
{\mathrm{sin}}^2(\frac{\pi\tau}{\beta})-R^2)}/\Lambda^2}}+1}
\def\m{e^{-{{(r^2 + \frac{\beta^2}{\pi^2}
{\mathrm{sin}}^2(\frac{\pi\tau}{\beta})-R^2)}/\Lambda^2}}}
\def\m1{e^{-{{(r^2 + \frac{\beta^2}{\pi^2}
{\mathrm{sin}}^2(\frac{\pi\tau}{\beta}))}/\Lambda^2}}}
\def\sin{{\mathrm{sin}}}
\def\cos{{\mathrm{cos}}}

\def\g{\gamma}
\def\d{\delta}
\def\l{\Lambda}
\def\vp{\phi}
\def\varphi{\phi}
\def\b{\beta}
\def\v0{\varphi_0}
\def\ka{\kappa}
\def\arg{\frac{2\pi\tau}{\b}}
\def\be{\begin{equation}}
\def\ee{\end{equation}}
\def\bea{\begin{eqnarray}}
\def\eea{\end{eqnarray}}
\documentclass[12pt]{article}
\title{Finite Temperature Phase Transition in $\vp^6$ potential}
\author{Hatem Widyan\thanks{E--mail : widyan@ahu.edu.jo} \\
    {Physics Department} \\
    {Al-Hussein Bin Talal University, Ma'an, Jordan}
    }

\begin {document}
\maketitle
\begin {abstract}

The temperature dependance of the action in the thin-wall and
thick-wall limits is obtained analytically for the $\phi^6$ scalar
potential. The nature of the phase transition  is investigated
from the quantum tunnelling regime at low temperatures to the
thermal hopping regime at high temperatures. It is first-order for
the case of a thin wall while for the thick wall it is second-
order.

\end{abstract}
\begin {section}*{\bf 1. Introduction}
%
%

 The existence of phase transition associated with spontaneous
symmetry breaking may appear during the evolution of the the
universe. Such a phase transition may influence to the large-scale
structure of the universe \cite{kirzhints}. Both the order of the
phase transition and its strength are basic ingredients for a
quantitative discussion of the transition at some energy scale. In
general, a symmetry-breaking phase transition can be first or
second order one. The decay of metastable state at a given
temperature $T$ can be written in the from $\Gamma=A{\rm
e}^{-S_E(T)}$, with $S_E(T)$ being the Euclidean action of the
saddle-point configuration  and $A$ being the prefactor determined
by the associated fluctuations. At zero temperature, the decay is
determined by quantum effects. With increasing temperature, the
nature of the decay changes from quantum to classical. The
function $S_E(T)$ might either be a smooth function of temperature
or exhibit a kink with a discontinuity in its derivative at some
temperature $T_c$. In the former case, the transition from the
quantum tunneling regime is said to be of second order while in
the latter case it is said to be of first order.

 We have considered in an earlier work the $\vp^4$ theory with
different  symmetry breaking terms \cite{Hatem,Hatem0}, where we
have obtained numerical as well as analytical solution for
different values of the asymmetric term. In this paper we extend
our recent work in $\vp^6$ potential \cite{Hatem1} which has been
investigated by many authors in the context of condensed matter as
well as particle physics (see for example
\cite{Bergner,Amaral,Flores,Joy,Zamo,Lu,Kim}). In \cite{Hatem1},
the scalar potential $\phi^6$ is studied at zero temperature and
at high temperature. The equations of motion are solved
numerically to obtain $O(4)$  spherical symmetric and $O(3)$
cylindrical bounce solutions. Also an analytical solution for the
bounce is presented in the thin wall-wall as well as thick-wall
limits, where the potential is given by

\begin{equation}
 U(\varphi) = g \, \varphi^2 (\varphi^2-\varphi_0^2)^2-\delta
 \varphi^2,
 \label{bergner2}
\end{equation}
where $\varphi_0^2=-\lambda/{2g}$ and
$\delta=(\lambda^2/g-2m^2)/4$. Fixing the value of $\phi_0=2.39$
and $g=0.07$, the only adjustable parameter in the potential is
$\delta$ and its range is $0<\delta<g\phi_0^4=2.28$. In this paper
we extend our calculation of the action in \cite{Hatem1} to finite
temperatures and study the nature of the transition. We propose a
general ansatz at finite temperature in the thin wall ($\delta
\rightarrow 0$) and thick wall ($\delta \rightarrow 2.28$) limits.
We find that for a thin wall the transition is first-order while
for a thick wall it is second-order. In section $2$, we present
our analytical calculations of the action at finite temperature in
the thin wall limit, while the calculations for the thick wall
limit are presented in section $3$. Section $4$ contains our
conclusions. The algebraic expressions for integrals appearing in
the analytical formalism are given in Appendixes A and B.


\end {section}
\begin {section}*{\bf 2. Action at finite temperature in the thin wall
limit}
%
%

The action at finite temperature of a single scalar field $\vp$ is
given by the following formula
\be S(T)= 4\pi \int_{-\b/2}^{\b/2} d\tau \int_0^{\infty} dr r^2
\Bigg[\frac{1}{2} \Bigg(\frac{\partial \vp}{\partial\tau}\Bigg)^2
+ \frac{1}{2} \Bigg(\frac{\partial \vp}{\partial r}\Bigg)^2 +
U(\vp) \Bigg] ~. \label{action} \ee
The equation of motion derived from the above action is given by
the following expression
\be {\partial^2\vp \over \partial\tau^2} + {\partial^2\vp \over
\partial r^2} + {2 \over r }{\partial\vp \over \partial r} =
{\partial U(\vp,T) \over \partial\vp} ~, \label{eom} \ee
with boundary conditions
\be \vp \to \vp_- \quad {\rm as} \quad r \to \infty ,\quad
\partial\vp/\partial\tau = 0 \quad {\rm at} \quad \tau = \pm \beta/2,0
~, \label{bcs} \ee
where $\vp_-$ is the false vacuum of the potential $U$, $\beta$ is
the period of the solution and $r=\sqrt{\vec{x}^2}.$

Following \cite{Hatem1,Bergner,Flores,Joy} we assume for the
solution of the equation of motion the following ansatz:

\be \vp^2(r,\tau)=\frac{\g}{\t} ~, \label{ansatz} \ee
which is periodic in the interval $(-\b/2,\b/2)$ and satisfies the
required boundary conditions (Eq.~(\ref{bcs})), viz
\be \frac{\partial\vp}{\partial r}=0 ~\mathrm{at}~ r=0,
~\frac{\partial\vp}{\partial\tau}=0 ~\mathrm{at}~ \tau=0
~\mathrm{and} ~ \pm\b/2, ~\mathrm{and} ~\vp=0 ~\mathrm{as}~ r \to
\infty ~. \label{bcs1}\ee
We evaluate the action for potential given by Eq.~(\ref{bergner2})
\be
 U(\varphi) = g \, \varphi^2 (\varphi^2-\varphi_0^2)^2-\delta
 \varphi^2.
 \label{pot}
\ee
After substituting the ansatz function Eq.~(\ref{ansatz}) into the
equation of motion Eq.~(\ref{eom}), we have
\bea \frac{\sqrt\g}{(\t)^{5/2}} \Bigg[ \frac{3
r^2}{\l^4}+\frac{3\b^2}{4\pi^2 \l^4} \sin^2(\arg) \Bigg] \nonumber
\\  [0.3cm]
+ \frac{\sqrt\g}{(\t)^{3/2}} \Bigg[ -\frac{4
r^2}{\l^4}+\frac{3}{\l^2}+\frac{1}{\l^2} \cos(\arg) -
\frac{\b^2}{\pi^2 \l^4} \sin^2(\arg) \Bigg] \nonumber \\  [0.3cm]
+ \frac{\sqrt\g}{(\t)^{1/2}} \Bigg[ \frac{ r^2}{\l^4} -
\frac{3}{\l^2} + \frac{\b^2}{4\pi^2 \l^4} \sin^2(\arg) -
\frac{1}{\l^2} \cos(\arg) \Bigg] \nonumber \\  [0.3cm]
= \frac{6g \g^{5/2}}{(\t)^{5/2}} -8 g \v0^2
\frac{\g^{3/2}}{(\t)^{3/2}} \nonumber \\ [0.3cm]
+ 2 (g \v0^4-\d) \frac{\g^{1/2} }{(\t)^{1/2}} ~. \label{twa} \eea
In the thin wall limit, the bounce solution is constant except in
a narrow region near the wall. Hence, by equating terms with
different powers of exponentials separately in Eq.~(\ref{twa}), we
have with $r^2+\frac{\b^2}{4\pi^2} \sin^2(\arg)\approx R^2$,
\bea 2 g \g^2
=\frac{R^2}{\l^4} \Bigg[ 1-a\frac{\l^2}{R^2} \Bigg]
~. \nonumber \\ [0.3cm]
2 g \v0^2 \g=\frac{R^2}{\l^4} \Bigg[ 1 - b \frac{\l^2}{R^2} \Bigg]
~. \nonumber \\ [0.3cm]
2(g \v0^4 -\d)= \frac{R^2}{\l^4} \Bigg[ 1 -d \frac{\l^2}{R^2}
\Bigg] ~. \label{abd} \eea
The parameters $a$, $b$ and $d$ are found by the requirement that
the variation of $S(T)$ with respect to the parameters $R$, $\l$
and $\g$ in Eq.~(\ref{ansatz}) vanishes.

The integrals in the action are obtained in powers of $\l^2 /R^2$
using the usual methods for evaluating integrals of the Fermi
function (see eg. Huang \cite{huang}). We get
\bea
S(T)  &=&  4\pi \int_{-\b/2}^{\b/2} d\tau \int_0^{\infty} dr r^2
\Bigg[\frac{1}{2} \Bigg(\frac{\partial \vp}{\partial\tau}\Bigg)^2
+ \frac{1}{2} \Bigg(\frac{\partial \vp}{\partial r}\Bigg)^2 + g
\vp^6 -2g \v0^2 \vp^4 +(g \v0^4-\d) \vp^2 \Bigg]
\nonumber \\
%
%
 & = & 4\pi \ka R^4  E_3 \, \g \Bigg[
\frac{\ka^2}{4\l^2} \Bigg(\frac{E_{1T}}{E_3} +\frac{1}{2}
 \frac{E_{0T}}{E_3} \frac{\l^2}{R^2}\Bigg)+ \frac{1}{4\l^2}
\Bigg(1+\frac{3}{2} \frac{E_1}{E_3}\frac{\l^2}{R^2}+
\frac{\pi^2}{8}\frac{E_0}{E_3} \frac{\l^4}{R^4}\Bigg) \nonumber
\\ [0.3cm]
& + & \frac{g \g^2}{2} \Bigg(\frac{4}{3} - \frac{3E_1}{E_3}
\frac{\l^2}{R^2} + \Big(\frac{\pi^2}{6}+\frac{1}{2}\Big)
\frac{E_0}{E_3} \frac{\l^4}{R^4}\Bigg) \nonumber \\ [0.3cm]
& - &2 g \v0^2 \g \Bigg(\frac{2}{3} - \frac{E_1}{E_3}
\frac{\l^2}{R^2} + \frac{\pi^2}{12}
\frac{E_0}{E_3}\frac{\l^4}{R^4}\Bigg) \nonumber \\ [0.3cm]
& + & \frac{1}{2}(g \v0^4-\d) \Bigg(\frac{4}{3} +
\frac{\pi^2}{6}\frac{E_0}{E_3} \frac{\l^4}{R^4} \Bigg) \Bigg] ~,
\label{actionint} \eea
where
\bea E_0(\ka^2) & = & \int_0^1 \frac{dt}{\sqrt{1-t^2}\sqrt{1-\ka^2
t^2}} ~, \nonumber \\ [0.3cm]
E_1(\ka^2) & = & \int_0^1 \frac{dt\sqrt{1-\ka^2
t^2}}{\sqrt{1-t^2}} ~, \nonumber \\ [0.3cm]
E_3(\ka^2) & = & \int_0^1 \frac{dt(1-\ka^2
t^2)^{3/2}}{\sqrt{1-t^2}} ~, \nonumber \\ [0.3cm]
E_{0T}(\ka^2) & = & \int_0^1 \frac{dt (1-t^2) t^2} {\sqrt{1-
t^2}{\sqrt{1-\ka^2 t^2}}} \nonumber  ~, \\ [0.3cm]
E_{1T}(\ka^2) & = & \int_0^1 \frac{dt (1-t^2) t^2 \sqrt{1-\ka^2
t^2}}{\sqrt{1-t^2}} \nonumber ~, \eea
are the complete elliptic integrals and they can be represented in
terms of the basic complete elliptic integrals $E_0$ and $E_1$
(see Appendix \ref{app1}), $\ka=\frac{\b}{\pi R}$ and
$t=\sin\frac{\pi}{\b}\tau$.

We now determine the parameters $a$, $b$ and $d$ by demanding the
vanishing of $dS(T)/dR^2$, $dS(T)/d\l^2$ and $dS(T)/d\g$.
Differentiating Eq.~(\ref{actionint}) and using Eq.~(\ref{abd}),
we find that to leading order in $\l^2/R^2$,
\bea -4 a+ 8 b -4 d +3 +\frac{3(E_1-2\ka^2 E_1')}{3E_3-2\ka^2
E_3'} +
\frac{3 \ka^2 (E_{1T} -2 \ka^2 E_{1T}')}{3E_3-2\ka^2 E_3'} = 0 ~, \nonumber \\
[0.3cm]
a \Bigg(\frac{3}{4} \lambda - \frac{1}{2} g  \g^2 \epsilon_T
\frac{E_3}{E_1}\Big(1-\frac{\lambda}{2g\g^2}\Big) \Bigg) + b
\Bigg( \frac{1}{2} g \g^2 \epsilon_T \frac{E_3}{E_1} -\lambda
\Bigg) + \frac{1}{4} \frac{E_0}{E_1} \lambda = 0 ~, \nonumber
\\ [0.3cm]
-a+\frac{4b}{3}-\frac{1}{3} d-\frac{1}{4} \epsilon_T = 0 ~,
\label{derivative} \eea
where $\lambda =2 g \g^2-2 g \v0^2 \g$, $E_1'$ is the derivative
of $E_1$ with respect to $\ka^2$ (and similarly for $E_{1T}'$ and
$E_3'$), and $\epsilon_T=E_1/E_3-\ka^2 E_{1T}/E_3 -1$.

Note that in the limit of zero temperature ($\kappa \rightarrow
\infty$), Eq.~(\ref{derivative}) reduces to
$$3-2a+4b-2d=0, ~~~~~~ 2+3a-4b=0, ~~~~~~~~{\rm and} ~~~~~~~~~ -3a+4b-d=0,$$
respectively which are obtained earlier  \cite{Hatem1}. Also, in
the limit of high temperature (i.e. $\kappa \rightarrow 0$), they
reduce to
$$1-a+2b-d=0, ~~~~~~~~~ 1+3a-4b=0, ~~~~~~{\rm and} ~~~~~~~~~
-3a+4b+d=0,$$ respectively which are also obtained earlier
\cite{Hatem1}.

By using Eq.~(\ref{derivative}), we can find a relation between
the constants $a$, $b$ and $d$,
\bea d-a=\frac{3}{4}\epsilon_T +\frac{3}{2}+\frac{c}{2}~, \nonumber \\
[0.3cm]
d-b=\frac{3}{8}(\epsilon_T +c+3) ~, \nonumber \\ [0.3cm]
b-a=\frac{3}{8}(1+\epsilon_T)+\frac{c}{8} ~, \eea
where $c$ is given by the following expression
\be c=\frac{3(E_1-2\ka^2 E_1')}{3E_3-2\ka^2 E_3'} +
\frac{3\ka^2(E_{1T}-2\ka^2E_{1T}')}{3E_3-2\ka^2E_3'} ~. \nonumber
\ee
Thus Eq.~(\ref{actionint}) can be expressed in terms of $a$, $b$
and $d$. It reads as
\bea S(T)&=&\frac{2 \pi \ka}{g \v0^2} E_3
\Bigg(\frac{R}{\l}\Bigg)^6 \Bigg[\Bigg(1-b\frac{\l^2}{R^2} \Bigg)
\Bigg(\frac{1}{2}\frac{E_1}{E_3}-\frac{2}{3}(b-a)\Bigg) +
\frac{E_1}{E_3}\Bigg( \frac{3}{32} \Big(2 b -d -3 -3 \epsilon_T
\Big) \nonumber \\&+&   \frac{3}{8}\frac{E_1}{E_3}+\frac{1}{8}
\frac{E_0}{E_3} + \frac{1}{8} \frac{E_{0T}}{E_3} \Bigg)
\frac{\l^2}{R^2} \Bigg] ~. \label{actionf} \eea
With these expressions we can calculate the values of $\g$, $R$
and $\l$ and also for the action $S(T)$. In calculating $S(T)$ for
$\b \to \infty$, the integrals $E_{i4}$ are to be used (see
Appendix \ref{app1}). In these cases $\ka >1$ and restrict the
upper limit of the elliptic integral to $1/\ka$ as they become
complex for larger values of $\ka$.

%
%

As was noticed in  \cite{Hatem0}, there is a singularity in
($b-a$) due to $E_0(\kappa)$ becoming singular at $\kappa=1$. The
values of $S(T)$ are obtained and plotted for $\delta=0.1$ and
$\delta=0.3$ as shown in Figs. $1$ and $2$ respectively. The
inverse of the temperature $\beta_\star$ is defined by
\cite{Hatem} $\beta_\star=S_4/S_3$. The transition point $\beta_c$
can in principle be different from $\beta_\star$, but in the TWA
are equal \cite{Hatem, Hatem0}. In our results (Figs. 1 and 2) we
can determine $\beta_\star$ by extending the horizontal part of
the curve to the left. For Fig. 1, for example, this yields
$\beta_\star=$, which is close to the value of $\beta_c$ obtained
numerically and analytically \cite{Hatem0}. We conclude that the
singularity is an artifact of the method, and does not represent
the transition point. The phase transition actually takes place at
a much lower value of $\beta_c$, and is first-order.

It can be shown that in the limit of zero temperature ($\kappa
\rightarrow \infty$) and in the limit of high temperature ($\kappa
\rightarrow 0$), the action in Eq.~(\ref{actionf}) reduces to the
action given earlier \cite{Hatem1}.

\end {section}

\begin {section}*{\bf 3. Action at finite temperature in the thick wall
limit}

The form of the bounce in Eq.~(\ref{ansatz}) suggests that the
thick wall limit, which would correspond to small values of
$R^2/\l^2$, would be obtained by approximating the Fermi function
by the Maxwell-Boltzmann function, which leads to a Gaussian:

\be \vp^2(r,\tau)={\g}\,{\m1} ~, \label{ansatz1} \ee
which satisfies the boundary conditions given by Eq.~(\ref{bcs1}).
The action for this form of bounce is found to be

\bea S(T) & = & \frac{\pi}{2} \g \Gamma(\frac{3}{2}) \l^4 {\rm
e}^{-x^2/2} x  I_0(\frac{x^2}{2})  \Bigg[ \frac{3}{4\l^2} +
\frac{1}{4\l^2}\frac{I_1(\frac{x^2}{2})}{I_0(\frac{x^2}{2})}-2g\phi_0^2
\g (\frac{1}{2})^{3/2} {\rm e}^{-x^2/2}
\frac{I_0(x^2)}{I_0(\frac{x^2}{2})} \nonumber
\\ [0.3cm]
& + & g \g^2 (\frac{1}{3})^{3/2} {\rm e}^{-x^2}
\frac{I_0(\frac{3x^2}{2})}{I_0(\frac{x^2}{2})}  +  (g \v0^4-\d)
\Bigg] ~, \label{actionint1}
\eea
where $x=\beta/{\pi\l}$ and$I_\nu(x^2)$ are the modified Bessel
functions.

Equation (\ref{abd}) then reduces to

\be 2g\g^2=-\frac{a}{\l^2}, ~~~~ 2\phi_0^2\g=-\frac{b}{\l^2}, ~~~~
2(g\phi_0^4-\delta)=-\frac{d}{\l^2}. \ee
Here we assume $\gamma^2 <<1$, hence $a=0$. The values of $b$ and
$d$ are obtained by demanding $dS(T)/d\g=dS(T)/d\l=0$. This gives
the following:

\bea
\frac{3}{4}+\frac{1}{4}\frac{I_1(\frac{x^2}{2})}{I_0(\frac{x^2}{2})}
+\frac{b}{\sqrt 2} {\rm e}^{-x^2/2}
\frac{I_0(x^2)}{I_0(\frac{x^2}{2})}-\frac{d}{2}=0  \label{dstg}
\nonumber \\ %
\frac{3}{8} {\rm e}^{-x^2/2}\Bigg(
{I_0(\frac{x^2}{2})}+{I_1(\frac{x^2}{2})}\Bigg)&+&\frac{x^2}{4}
{\rm e}^{-x^2/2}
\Bigg({I_0(\frac{x^2}{2})}-{I_1(\frac{x^2}{2})}\Bigg) + b
(\frac{1}{2})^{5/2} {\rm e}^{-x^2} I_0(x^2) F \nonumber \\ &-&
\frac{d}{8} {\rm e}^{-x^2} {I_0(\frac{x^2}{2})} E=0 \label{dstl}
\eea
where

\bea E=6+2 x^2
\Big(1-\frac{I_1(\frac{x^2}{2})}{I_0(\frac{x^2}{2})}  \Big)  \nonumber \\
F=3+2 x^2 \Big(1-\frac{I_1(x^2)}{I_0(x^2)} \Big). \eea

In order to check our results, note that in the limit of zero
temperature ($x \rightarrow \infty$), Eqs.~(\ref{dstg}) and
(\ref{dstl}) reduce to
$$2+b-d=0 ~~~~~~{\rm and} ~~~~~~~~~ 2+b-2d=0$$ respectively which are
obtained earlier  \cite{Hatem1}. Also, in the limit of high
temperature (i.e. $x \rightarrow 0$), they reduce to
$$\frac{3}{16}+\frac{b}{2^{5/2}}-\frac{d}{8}=0 ~~~~~~{\rm and} ~~~~~~~~~
\frac{3}{8}+(\frac{1}{2})^{3/2}\frac{3b}{2}-\frac{3d}{4}=0$$
respectively which are obtained earlier  \cite{Hatem1}.

Using the Eqs.~(\ref{dstg}) and (\ref{dstl}), the values of $b$
and $d$ are given by

\be b=\frac{\frac{E}{16} \Big(3 I_0(\frac{x^2}{2})+
I_1(\frac{x^2}{2})\Big)-\frac{3}{8} \Big(I_0(\frac{x^2}{2})+
I_1(\frac{x^2}{2})\Big) -   \frac{x^2}{4} \Big(I_0(\frac{x^2}{2})-
I_1(\frac{x^2}{2})\Big) }{(\frac{1}{2})^{5/2}(F-E) I_0(x^2){\rm
e}^{{-x^2}/2}} \ee

\be d=\frac{F \Big(\frac{3}{2} I_0(\frac{x^2}{2})+ \frac{1}{2}
I_1(\frac{x^2}{2})\Big)-3 \Big(I_0(\frac{x^2}{2})+
I_1(\frac{x^2}{2})\Big) -   2 x^2 \Big(I_0(\frac{x^2}{2})-
I_1(\frac{x^2}{2})\Big)}{(F-E) I_0(\frac{x^2}{2})} \ee
In the limit of zero temperature ($x \rightarrow \infty$), $b=-2$
and $d=0$, and in the limit of high temperature ($x \rightarrow
0$), $b=-\sqrt{2}$ and $d=-1/2$, which are the same values
obtained earlier \cite{Hatem1}.

Thus the action yields
\be S(T)=\frac{\pi^2}{g \phi_0^2}b \Gamma(\frac{3}{2}) x
I_0(\frac{x^2}{2}) {\rm e}^{{-x^2}/2}
\Big(\frac{3}{4}+\frac{1}{4}\frac{I_1(\frac{x^2}{2})}{I_0(\frac{x^2}{2})}\Big)+
b (\frac{1}{2})^{3/2} {\rm e}^{{-x^2}/2} \frac{
I_0(x^2)}{I_0(\frac{x^2}{2})}-\frac{d}{2}. \ee

It can be shown that in the limit of zero temperature ($x
\rightarrow \infty$) and in the limit of high temperature ($x
\rightarrow 0$), the action obtained the the last equation reduces
to the action given earlier \cite{Hatem1}.

For a given value of temperature (i.e. $x$), we can calculate $b$
and $d$. Here, $\gamma$ and $\Lambda$ are determined. Thus we can
calculate the action at different values of temperatures. Fig. 3
shows the value of the action at different values of inverse of
temperature for $\delta=2.0$. As we can see from the figure, the
action goes smoothly from the zero temperature regime to the high
temperature regime without any singularity at the transition
point. This means that in the thick-wall limit the transition is
second order. Moreover, the action at zero temperature is
independent of the value of $\delta$ as shown in our earlier work
\cite{Hatem1}, which has been also verified in our calculations
here.

\end {section}

\begin {section}*{\bf 4. Conclusions}

We now discuss the nature of the transition as we go from zero to
high temperatures. In quantum mechanics, definitive criteria for
the continuity or discontinuity (corresponding to second order and
first order respectively) in the derivative of the action have
been obtained by Chudnovsky \cite{chudnovsky} and Garriga
\cite{garriga}. It has even been shown that the lowest action at
any temperature is possessed by either the zero temperature or the
high temperature solutions.

 In quantum field theory the situation seems to be
different. Both Ferrera \cite{ferrera} and we \cite{Hatem} find
that there is an interpolating solution which can be used to
determine whether transition is first order or second order (i.e.
with or without a kink).

We have found that for a thin wall ($\delta \rightarrow 0$) the
interpolating solution has a singularity at $\beta=\pi R$. But it
is not a real singularity at this point. It is due to the
expansion method used in the calculations. Our numerical solutions
show a kink is present in the TWA, showing that the transition is
first order. However, for $\delta=2$ (thick wall), we find there
is no kink and the transition is smooth (second order).

We would to mention here that the Eqs.~(\ref{ansatz}) and
(\ref{ansatz1}) are not an exact solutions of the equations of
motion at finite temperature although they satisfy the required
boundary conditions. We think they represent a reasonable
approximation because in the case of thin wall approximation at
zero temperature Eq.~(\ref{ansatz}) is a solution of the equation
of the motion, see \cite{Flores}.

 It is suggested that our method could be used to study in detail the
nature of the phase transition in electroweak theory. Such a study
could be of importance in models of electroweak baryogenesis and
other phenomena in the early universe.

\end {section}

\bibliography{plain}

\begin{appendix}
%
\section{EXPRESSIONS OF ELLIPTIC INTEGRAL IN TERMS OF THE BASIC
INTEGRALS $E_0$ AND $E_1$} \label{app1}
For $\ka <1$
\bea
E_0 & = & \int_0^1 \frac{dt}{\sqrt{1-t^2}\sqrt{1-\ka^2 t^2}}
\\ [0.3cm]
E_1 & = & \int_0^1 \frac{dt\sqrt{1-\ka^2 t^2}}{\sqrt{1-t^2}}
\\ [0.3cm]
E_3 & = & \int_0^1 \frac{dt(1-\ka^2 t^2)^{3/2}}{\sqrt{1-t^2}}
=E_1(\frac{4}{3}-\frac{2}{3}\ka^2) + E_0(\frac{\ka^2-1}{3})
\\ [0.3cm]
E_{0T} & = & \int_0^1 \frac{dt (1-t^2)
t^2}{\sqrt{1-t^2}\sqrt{1-\ka^2}} = E_1(\frac{2}{3}\frac{1}{\ka^4}
- \frac{1}{3} \frac{1}{\ka^2}+ E_0 (\frac{2}{3}\frac{1}{\ka^2}-
\frac{2}{3} \frac{1}{\ka^2})\\
[0.3cm]
E_{1T} & = & \int_0^1 \frac{dt\sqrt{1-\ka^2 t^2} t^2
(1-t^2)}{\sqrt{1-t^2}} =
\frac{2E_1}{15}(\frac{1}{\ka^4}-\frac{1}{\ka^2}+1) +
\frac{E_0}{15}(-\frac{2}{\ka^4}+\frac{3}{\ka^2}-1)
\\ [0.3cm]
E_1'& = &\frac{E_1-E_0}{2\ka^2}
 \\[0.3cm]
E_3' & = & \frac{E_0}{2}(1-\frac{1}{\ka^2}) + E_1(-1+\frac{1}{2\ka^2})
 \\ [0.3cm]
E_{1T}' & = &
\frac{E_1}{15}(-\frac{4}{\ka^6}+\frac{3}{2\ka^4}+\frac{1}{\ka^2})
+
\frac{E_0}{15}(\frac{4}{\ka^6}-\frac{7}{2\ka^4}-\frac{1}{2\ka^2})
\eea
For $\ka > 1$
\bea
E_0(1/\ka^2) & =&  \int_0^1
\frac{dt}{\sqrt{1-t^2}\sqrt{1-t^2/\ka^2}}
 \\ [0.3cm]
E_1(1/\ka^2) &  = & \int_0^1 \frac{dt\sqrt{1-t^2/\ka^2}}
{\sqrt{1-t^2}}
 \\ [0.3cm]
E_{04}(\ka^2) & =&  \int_0^{1/\ka} \frac{dt}{\sqrt{1-t^2}\sqrt{1-\ka^2
t^2}}
\\ [0.3cm]
E_{14}(\ka^2)&  = & \int_0^{1/\ka} \frac{dt\sqrt{1-\ka^2
t^2}}{\sqrt{1-t^2}} = \ka E_1(1/\ka^2)-\frac{\ka^2-1}{\ka}
E_0(1/\ka^2)
\\ [0.3cm]
E_{34}(\ka^2) & = & \int_0^{1/\ka}
\frac{dt(1-\ka^2t^2)^{3/2}}{\sqrt{1-t^2}}
=\frac{1}{\ka}\Bigg[E_1(1/\ka^2)\Big(\frac{4\ka^2}{3}-
\frac{2\ka^4}{3}\Big)
\nonumber \\
& + & E_0(1/\ka^2)\Big(1-\frac{5\ka^2}{3}
 +  \frac{2\ka^4}{3}\Big) \Bigg]
\\ [0.3cm]
E_{0T4} (\ka^2) &=& \int_0^{1/\ka}  \frac{dt\sqrt{1-\ka^2 t^2}
t^2} {(1-t^2)\sqrt{1-t^2}} = \frac{1}{\ka} \Bigg[ E_1(1/\ka^2)
\Big(-\frac{1}{3}+\frac{2}{3\ka^2} \Big )
\nonumber \\
& + & E_0(1/\ka^2) \Big (\frac{1}{3}-\frac{1}{3\ka^2} \Big) \Bigg]
\\ [0.3cm]
E_{1T4} (\ka^2)& =&  \int_0^{1/\ka}  \frac{dt\sqrt{1-\ka^2 t^2}
t^2 (1-t^2)}{\sqrt{1-t^2}} = \frac{1}{\ka} \Bigg[ E_1(1/\ka^2)
\Big(-\frac{2}{15}+\frac{2}{15\ka^2}+\frac{2\ka^2}{15} \Big )
\nonumber \\
& + & E_0(1/\ka^2) \Big
(\frac{1}{5}-\frac{1}{15\ka^2}-\frac{2\ka^2}{15}
\Big) \Bigg]
\\ [0.3cm]
\frac{dE_0(1/\ka^2)}{d\ka^2} & = & \frac{1}{2\ka^2}E_0(1/\ka^2) -
\frac{E_1(1/\ka^2)}{2(\ka^2-1)}
\\ [0.3cm]
\frac{dE_1(1/\ka^2)}{d\ka^2} & = &
\frac{1}{2\ka^2}\Bigg(E_0(1/\ka^2)-E_1(1/\ka^2)\Bigg)
\\ [0.3cm]
\frac{dE_{14}(\ka^2)}{d\ka^2} & = &
\frac{1}{2\ka}\Bigg(E_1(1/\ka^2)-E_0(1/\ka^2)\Bigg)
\\ [0.3cm]
\frac{dE_{34}(\ka^2)}{d\ka^2} & = &
E_(1/\ka^2)\Bigg(\frac{1}{2\ka}-\ka\Bigg)+
E_0(1/\ka^2)\Bigg(-\frac{1}{\ka}+k\Bigg)
\\ [0.3cm]
\frac{dE_{0 T4}(\ka^2)}{d\ka^2} & = & E_0(1/\ka^2)
\Bigg(-\frac{1}{6\ka^3}+\frac{2}{3\ka^5} \Bigg) +
E_1(1/\ka^2)\Bigg(\frac{1}{6\ka^3}-\frac{4}{3\ka^5}\Bigg)
\\ [0.3cm]
 \frac{dE_{1T4}(\ka^2)}{d\ka^2} & = & E_0(1/\ka^2)
\Bigg(\frac{2}{15\ka^5}-\frac{1}{15\ka^3}-\frac{1}{15\ka} \Bigg) +
E_1(1/\ka^2)\Bigg(-\frac{4}{15\ka^5}+\frac{1}{10\ka^3}+\frac{1}{15\ka}
\Bigg) \eea
%
%


\end{appendix}

\begin{figure}[ht]
\vskip 15truecm

\includegraphics{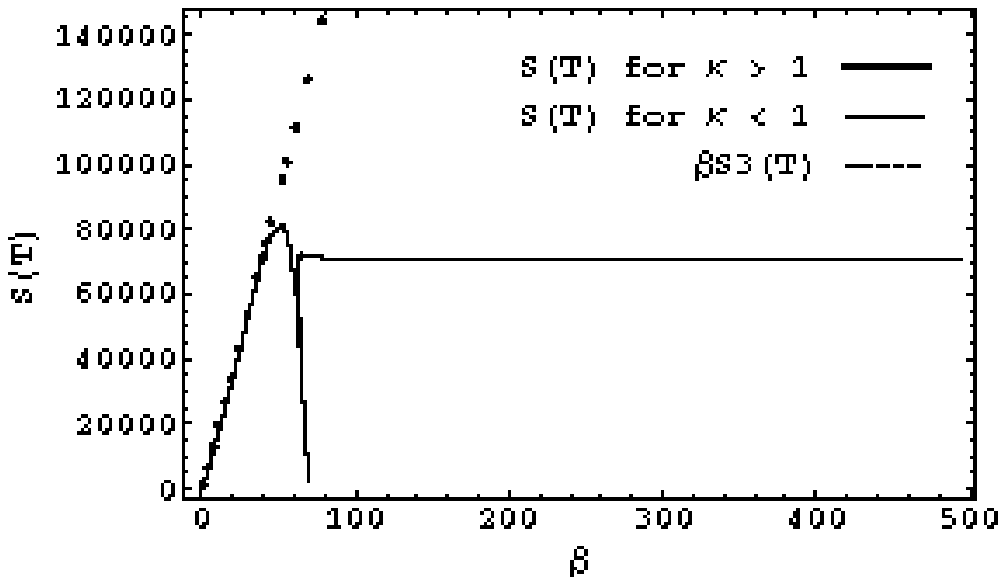}

\caption{Temperature dependence of the Euclidean actin in thin-
wall limit: $S(T)$ vs $\beta$ for $\delta=0.1$}
\end{figure}

\begin{figure}[ht]
\vskip 15truecm

\includegraphics{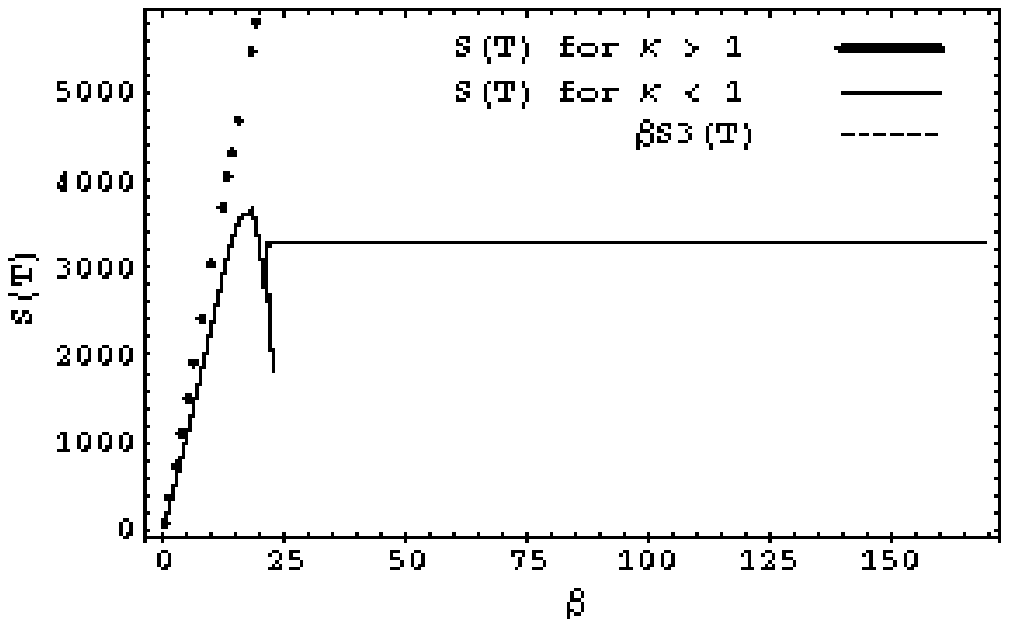}

\caption{Temperature dependence of the Euclidean actin in thin-
wall limit: $S(T)$ vs $\beta$ for $\delta=0.3$}
\end{figure}

\begin{figure}[ht]
\vskip 15truecm

\includegraphics{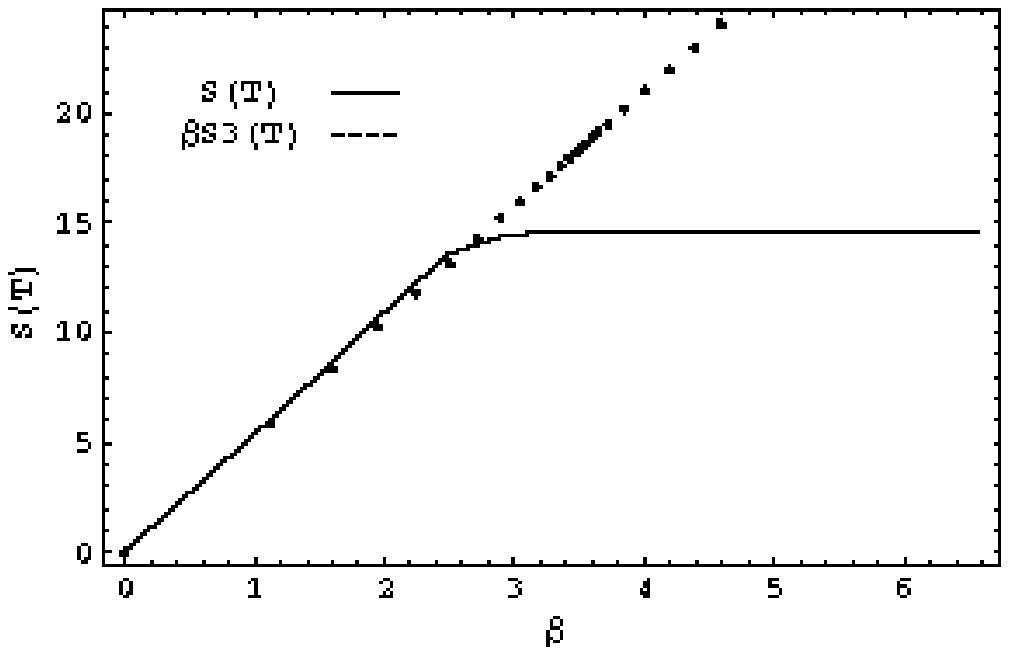}

\caption{Temperature dependence of the Euclidean actin in thick-
wall limit: $S(T)$ vs $\beta$ for $\delta=2.0$}

\end{figure}

\end{document}